\documentstyle{article}

\begin{document}
\title{Higher-order corrections to the two and three-fermion Salpeter equations.} 
\author{ J. Bijtebier\thanks{Senior Research Associate at the
 Fund for Scientific Research (Belgium).}\\
 Theoretische Natuurkunde, Vrije Universiteit Brussel,\\
 Pleinlaan 2, B1050 Brussel, Belgium.\\ Email: jbijtebi@vub.ac.be}
%\date{}
\maketitle
\begin{abstract}  
 We compare two opposite ways of performing a 3D reduction of the two-fermion Bethe-Salpeter equation beyond the instantaneous
approximation and Salpeter's equation. The  more usual method consists in performing an expansion around an instantaneous
approximation of the product of the free propagators (propagator-approximated reduction). The second method starts with an
instantaneous approximation of the Bethe-Salpeter kernel (kernel-approximated reduction). In both reductions the final 3D potential
can be obtained by following simple modified Feynman rules. The kernel-approximated reduction, however, does not give the correct
scattering amplitudes, and must thus be limited to the computation of bound states.\par
Our 3D reduction of the three-fermion Bethe-Salpeter equation is inspired by these results. We expand this equation
around positive-energy instantaneous approximations of the three two-body kernels, but these starting approximations are given by
propagator-approximated reductions at the two-body level. The three-fermion Bethe-Salpeter equation is first transformed into a set
of three coupled equations for three wave functions depending each on one two-fermion total energy, then into a set of three 3D
equations and finally into a single 3D equation. This last equation is rather complicated, as it combines
three series expansions, but we use it to write a manageable expression of the first-order corrections to the energy spectrum.   
\end{abstract}\newpage\noindent
equations. \newline
\tableofcontents

\section{Introduction}
The Bethe-Salpeter equation \cite{1,2}  is the usual tool for the study of relativistic bound states. The principal
difficulty in the treatment of this equation comes from the existence of unphysical relative time variables. In the two-body
Bethe-Salpeter equation the simplest way of eliminating the (in this case single) relative time variable consists in performing an
instantaneous approximation of the Bethe-Salpeter kernel (kernel-approximated reduction): one replaces the complicated kernel given
by field theory by a simpler expression, independent of the initial and final relative energies in the momentum space representation.
A 3D equation (Salpeter's equation \cite{3} ) follows immediately. It is also possible to get Salpeter's equation from an
instantaneous approximation of the product of the two free propagators: one replaces this product by a 3D propagator times a delta
(the "constraint") fixing the relative energy (propagator-approximated reduction). \par The derivation of Salpeter's equation using
a  kernel-approximated reduction appears to be the most natural one. For the calculation of the higher-order correction terms
(towards a theoretical equivalence with the original Bethe-Salpeter equation, at least for what concerns the bound state energy
spectrum), the  propagator-approximated reduction has been preferred by most authors [4-19]  including us \cite{17,19,18}. We found
an example of a kernel-approximated reduction in articles by Phillips and Wallace \cite{20}, who tune their instantaneous
approximation of the kernel in order to make the corrections vanish up to a given order. It is however our tentative of 3D reducing
the three-fermion Bethe-Salpeter equation \cite{21} which led us to consider this alternative approach (see below).\hfill\break
\par\noindent This work is divided into two parts. In the first part we present the two 3D reduction methods of the two-fermion
Bethe-Salpeter equation. We do not insist on the propagator-approximation reduction as we already closely examinated it elsewhere
\cite{17,19,18}, and we focus on the kernel-approximated reduction, comparing it with the other method.\par
 We meet a first difficulty coming from the
fact that the kernel-approximated reduction does not lead directly to an symmetric potential. In the case of a total-energy
depending potential like here, this does not necessarily imply that the final energy spectrum will not be real. It is however
interesting to get a 3D equation with a symmetric potential by performing a supplementary expansion. The resulting double expansion
(triple if we consider also the higher-order irreducible graphs contributing to the kernel) does not complicate the enumeration
of the contributions. It is indeed possible, as in the propagator-approximated reduction, to get the final potential by slightly
modifying the Feynman rules used in the calculation of the 4D off mass shell transition operator.\par
 The most concise way of
presenting the two reduction methods consists in writing their respective 3D off mass shell transition operators in terms of
the corresponding 4D off mass shell transition operator of field theory. Two reductions which share the same 3D off mass shell
transition operator correspond indeed to 3D equations which can be transformed into each other at the 3D level. In the
propagator-approximated reduction, the 3D operator is obtained from the 4D one by applying a constraint which is also satisfied by
the free solutions, so that the corresponding 3D equation gives not only the bound states, but also the correct physical scattering
amplitudes, although the original homogeneous Bethe-Salpeter equation was only valid for bound states. We loose this bonus with the
kernel-approximated reduction.\par We close this first part with the calculation of the second-order contributions to the 3D
potential using the kernel-approximated reduction method, in the case of a simple kernel given by a one-photon exchange graph (in
Feynman gauge). Each term is given by the residues of two poles in the initial and final relative-energy complex planes. The poles
can come from the propagators only, or one of them can come from the kernel. 
 The contributions of the kernel poles are not negligeable, at least in this case.
Beyond the ladder approximation, however, they could be partially cancelled by the contributions of the first crossed term.
\hfill\break\par\noindent In a recent work \cite{21}, we proposed a three-fermion Bethe-Salpeter equation with three
instantaneous, cluster-energy independent and positive-energy kernels. Such an equation can easily be transformed, without
supplementary approximation or perturbation expansion, into a 3D equation. The instantaneous two-body kernels were chosen in order to
reproduce the two-body potentials obtained with a propagator-approximated reduction in the pure two-body case. With this choice, our
3D equation satisfies a cluster separability requirement: "switching off" two of the three mutual interactions leads to a free
equation for one fermion, plus a 3D equation for the two other fermions, equivalent to the exact Bethe-Salpeter equation for
these two fermions. The remaining discrepancy (at least in the predicted energy spectrum) between the exact three-fermion
Bethe-Salpeter equation and our three-fermion 3D equation must then be considered as a three-body relativistic effect, the
computation of which is the purpose of the second part of the present work. Around a positive-energy instantaneous approximation of
the three two-body kernels (the three-body irreducible terms being entierely included in the perturbation), we perform three
consecutive expansions of the correction terms: the three-fermion Bethe-Salpeter equation is first transformed into a set
of three coupled equations for three wave functions depending each on one two-fermion total energy, then into a set of three 3D
equations and finally into a single 3D equation. The resulting expression for the 3D equation is rather complicated, as it combines
three series expansions, but we use it to write a manageable expression of the first-order corrections to the energy spectrum.  The
contributions of the poles of the propagators to this energy shift are easy to compute in terms of the Bethe-Salpeter kernels with
two of the three fermions on their mass shell, in the spirit of Gross' spectator model \cite{22,23}. The contributions of
the singularities of the kernels themselves (which could be important) remain to be computed but this should be done in the
framework of more specific models.\par What about cluster separability? We should not be surprised by uncorrect cluster-separated
limits, as the three-fermion homogeneous Bethe-Salpeter equation is valid only for three-body bound states. Correct limits would
again be a bonus. The cluster-separated limits of our 3D equation are two-fermion 3D equations built with the potentials
corresponding to the starting instantaneous two-body kernels. They are thus correct if these potentials are the full potentials
corresponding to reductions of the two-fermion Bethe-Salpeter equation. Another choice, like that based on the simpler Born
approximation, would lead to a 3D equation whith uncorrect cluster separated limits, but could still be used in the computation of
three-fermion bound states.         
 
\section{The two-fermion problem.}
\subsection{Notations.}
\noindent We shall write the Bethe-Salpeter equation for the bound states
 of two fermions \cite{1} as 
\begin{equation}\Phi = G^0 K \Phi,    \label{1}\end{equation}   where $\Phi$ is the Bethe-Salpeter  amplitude, function of the
positions $x_1,x_2$ or of the momenta 
$p_1,p_2$ of the fermions (according to the chosen representation). The operator $K$  is the Bethe-Salpeter kernel, given in a
non-local momentum representation by the sum of the irreducible two-fermion Feynman graphs. The operator $G^0$ is the free
propagator, given by the product
$G^0_1G^0_2$ of the two individual fermion propagators
\begin{equation} G^0_i = {1 \over p_{i0}-h_i+i\epsilon h_i}\,\beta_i = {p_{i0}+h_i\over p_i^2-m_i^2+i\epsilon}\,\beta_i 
\label{2}\end{equation}  where the $h_i$ are the Dirac free hamiltonians of eigenstates $\,\pm E_i:$
\begin{equation}h_i = \vec \alpha_i\, . \vec p_i + \beta_i\, m_i\qquad (i=1,2).  \label{3}\end{equation}
\begin{equation} E_i=\sqrt{h_i^2}=(\vec p_i^2+m_i^2)^{1\over 2}.   \label{4}\end{equation}    We shall define the
total (or external, CM, global) and relative (or internal) variables:
\begin{equation}X = {1 \over 2} (x_1 + x_2)\ , \qquad P = p_1 + p_2\ ,  \label{5}\end{equation} 
\begin{equation}x = x_1 - x_2\ , \qquad p = {1 \over 2} (p_1 - p_2). \label{6}\end{equation}  and give a name to the
corresponding combinations of the free hamiltonians:
\begin{equation}S = h_1 + h_2\ , \quad s = {1 \over 2} (h_1 - h_2).\label{7}\end{equation}
  
\subsection{3D reduction using an instantaneous approximation of the propagator.} The free
propagator $G^0$ will be approached by a carefully chosen expression
$G^\delta$, combining a constraint fixing the relative energy, and a global 3D propagator.
 There exists an infinity of
possible combinations [4-19]  which would be equivalent if all
correction terms could be computed. Our choice in this work will be
\begin{equation}G^\delta=\Lambda^{++}\int dp_0\,G^0(p_0)=-2i\pi \,\delta(p_0\!-\!s)\,\Lambda^{++}\,g^0\,\beta_1\beta_2.\label{8}\end{equation}  
where $\,\Lambda^{++}\,$ is the free positive-energy projector:
\begin{equation}\Lambda^{ij}=\Lambda_1^i\Lambda_2^j,
\qquad \Lambda_i^\pm={E_i\pm h_i\over 2E_i}. \label{9}\end{equation}
and $\,g_0\,$ the 3D propagator: 
\begin{equation}g^0\,\,={1 \over P_0-S + i\epsilon }
\label{10}\end{equation}
This choice of $\,G^{\delta}\,$ is not necessarily the best one for working in the pure two-fermion framework, but it is the
simplest one for the presentation of our two and three-fermion results. 
 We shall write the free propagator as the sum of the approached propagator, plus a remainder:
\begin{equation}G^0=G^{\delta}+G^R.  \label{11}\end{equation}  The Bethe-Salpeter equation  becomes then the inhomogeneous
equation
\begin{equation}\Phi=G^0K\Phi=(G^\delta +G^R)\,K\,\Phi=\Psi +G^RK\Phi, \label{12}\end{equation}  with
\begin{equation}\Psi=G^\delta K\Phi. \label{13}\end{equation}  Solving (formally) the
inhomogeneous equation (\ref{12}) and putting the result into (\ref{13}), we get
\begin{equation}\Psi=G^\delta K(1-G^RK)^{-1}\Psi=G^\delta K^T\Psi   \label{14}\end{equation}  where
\begin{equation}K^T=K(1-G^RK)^{-1}=K+KG^RK+...=(1-KG^R)^{-1}K  \label{15}\end{equation}  obeys
\begin{equation}K^T=K+KG^RK^T=K+K^TG^RK. \label{16}\end{equation}  The reduction series (\ref{15}) re-introduces in fact the
reducible graphs into the Bethe-Salpeter kernel, but with $G^0$ replaced by $G^R$. Equation (\ref{14}) has the same bound
state spectrum as the original Bethe-Salpeter equation.  \par
In the inversion of $\,(1\!-\!G_RK)\,$ and similar operators below, we have to assume that eq. $\,(1\!-\!G_RK)\Phi\!=\!0\,$ has no
solution for the values of $\,P_0\,$ considered (i.e. the values belonging to the bound state energy spectrum of the Bethe-Salpeter
equation).   \par
 The relative energy dependence of eq. (\ref{14}) can be easily eliminated:
\begin{equation}\Psi=\delta(p_0\! -\! s)\,\psi \label{17}\end{equation} 
and $\,\psi\,$ obeys:
\begin{equation}\psi\,=\,g_0\,V^T\,\psi\label{18}\end{equation} where $\,V^T\,$ is proportional to the positive-energy part
of $\,K^T\,$ with the initial and final relative energies fixed to $\,s:\,$
\begin{equation}V^T\,=-2i\pi\,\Lambda^{++}\,\beta_1\beta_2K^T(s,s)\,\Lambda^{++}.\label{19}\end{equation}
In less compact but more precise notations:    
\begin{equation}\beta_1\beta_2K^T(s,s)\,\equiv\,\int dp_0' dp_0 \delta(p'_0\! -\!s)\beta_1\beta_2K^T(p_0',p_0)\delta(p_0\!
-\!s).\label{20}\end{equation}
Note that we write $\,(p'_0,p_0)\,$ but $\,(s,s),\,$ as we keep $\,s\,$ in operator form.\par   
These 3D reductions can also be described in terms of transition operators. The 4D transition operator is
\begin{equation}T\,=\,K\,+\,K\,G^0\,K\,+\,\cdots\label{21}\end{equation}
\noindent and $\,K^T\,$ can be obtained by keeping only the $\,G^R\,$ part of $\,G^0\,$ in it. We have also
$$T=K(1\!-\!G^0K)^{-1}=K(1\!-\!G^RK\!-\!G^{\delta}K)^{-1}=K(1\!-\!G^RK)^{-1}(1\!-\!G^{\delta}K(1\!-\!G^RK)^{-1})^{-1}$$
\begin{equation}=K^T(1\!-\!G^{\delta}K^T)^{-1}=K^T+K^TG^{\delta}K^T+\cdots\label{22}\end{equation} 
so that the 3D transition operator is given by
\begin{equation}V^T\,+\,V^T\,g^0\,V^T\,+\,\cdots=
\,-2i\pi\beta_1\beta_2\,T^G\label{23}\end{equation}
with
\begin{equation}T^G\,=\,K^{T++}(s,s)\,+\,K^{T++}(s,s)\,G^{\delta}\,K^{T++}(s,s)\,+\cdots=\,T^{++}(s,s)\label{24}\end{equation}
\begin{equation}K^{T++}(s,s)\,=\,\beta_1\beta_2\,\Lambda^{++}\,\beta_1\beta_2K^T(s,s)\,\Lambda^{++}
\,=\,{-1\over2i\pi}\,\beta_1\beta_2V^T\label{25}\end{equation}
\begin{equation}T^{++}(s,s)\,=\,\beta_1\beta_2\,\Lambda^{++}\,\beta_1\beta_2\,T(s,s)\,\Lambda^{++}
.\label{26}\end{equation} 
We see that the 3D transition operator is a constrained form of that of field theory. Both operators become equal to the physical
scattering amplitude when both fermions are on their positive-energy mass shells. Conversely:
$$T=K^T+K^TG^{\delta}K^T+K^TG^{\delta}(K^T+K^TG^{\delta}K^T+\cdots)G^{\delta}K^T$$
\begin{equation}=K^T+K^TG^{\delta}K^T+K^TG^{\delta}\,T^G\,G^{\delta}K^T.\label{26b}\end{equation} 
A Bethe-Salpeter equation leading directly to the same 3D reduction can be obtained by replacing  the kernel $\,K\,$ by the
instantaneous positive-energy kernel $\,K^{T++}(s,s).\,$
In this case the 4D transition operator $\,T\,$ becomes equal to
$\,T^{++}(s,s)=T^G.$\par\noindent
The first-order correction term to the energy is
\begin{equation} P_0-P_0^{(0)}\,=\,\,-2i\pi\,<\Lambda^{++}\,\beta_1\beta_2\,(KG^RK)(s,s)\,\Lambda^{++}>.
\label{27}\end{equation}
We must not forget that $\,P_0^{(0)},\,\,(KG^RK)(s,s)\,$ and the zero-order wave functions are in general total energy dependent, so
that (\ref{27}) is in fact a numerical equation in $\,P_0\,$ (or a matricial equation in case of degenerescence). If the $\,P_0\,$
dependence of the zero-order potential is of higher-order, one has only to take the $\,P_0\,$
dependence of $\,P_0^{(0)}\,$ into account.

\subsection{3D reduction using an instantaneous approximation of the kernel.}
We shall now consider an approximation of the Bethe-Salpeter kernel:
\begin{equation}K\,=\,K^0\,+\,K^R.\label{28}\end{equation}
The Bethe-Salpeter equation becomes
\begin{equation}\Phi=G^0K^0\Phi+G^0K^R\Phi\label{29}\end{equation}  
\begin{equation}\Phi=(1-G^0K^R)^{-1}G^0K^0\Phi=G^KK^0\Phi\label{30}\end{equation}
with
\begin{equation}G^K=\,G^0\,+\,G^0K^RG^0+G^0K^RG^0K^RG^0+\cdots\equiv\, G^0+G^{KR}\,.\label{31}\end{equation}
If we now specialize $\,K^0\,$ to an instantaneous positive-energy kernel ($K^0\,$ independent of $\,p_0\,$ and equal to
$\,\beta_1\beta_2\Lambda^{++}\beta_1\beta_2K^0\Lambda^{++}\,$), eq. (\ref{30})  leads to the 3D
equation
\begin{equation}\phi\,=\,(g^0+g^{KR}\,)\,V\,\phi\label{32}\end{equation}
with
\begin{equation}\phi=\Lambda^{++}\int dp_0\,\Phi(p_0),\qquad V=-2i\pi\,\beta_1\beta_2K^0,
\label{33}\end{equation}
\begin{equation}g^{KR}=-{1\over2i\pi}\,\Lambda^{++}\,\int
dp'_0\,dp_0\,G^{KR}(p'_0,p_0)\,\beta_1\beta_2\,\Lambda^{++}.\label{34}\end{equation}
The first-order energy shift is, with the replacement $\,g^{KR}V\phi\approx g^{KR}(P_0\!-\!S)\phi:$  
\begin{equation}P_0-P^{(0)}_0=-{1\over2i\pi}\,<\,(P^0\!-\!S)\int
dp'_0\,dp_0\,G^0(p'_0)K^R(p'_0,p_0)G^0(p_0)\,\beta_1\beta_2\,(P^0\!-\!S)\,>.\label{35}\end{equation}
Here again, all elements of (\ref{35}) are total energy-dependent. Although (\ref{35}) is symmetric, the perturbation
potential $\,(P_0\!-\!S)g^{KR}V\,$ of (\ref{32}) is not. This does not exclude the possibility of a real energy spectrum:
if we admit an energy dependence of the potential (which we can not avoid), we have also the possibility of performing
an infinity of rearrangements of the equation. We can for example render the potential symmetric by expanding eq.
(\ref{32}) in the same way as eq. (\ref{12}) of section (2.2). We can equivalently start with the corresponding 4D
equation (\ref{30}):
\begin{equation}\Phi=(G^0+G^{KR})\,K^0\Phi=\chi+G^{KR}\,K^0\Phi\label{36}\end{equation}
\begin{equation}\chi=G^0K^0\Phi\,=G^0K^0(1-G^{KR}\,K^0)^{-1}\,\chi=G^0K^K\chi\label{37}\end{equation}
where
\begin{equation}K^K\,=K^0\,+\,K^0G^{KR}K^0\,+\,\cdots\label{38}\end{equation}
 is a symmetric positive-energy instantaneous potential, since it begins and ends with $\,K^0\,$. The 3D equation is then
\begin{equation}\eta\,=\,g^0\,V^K\,\eta,\qquad V^K\,=\,-2i\pi\,\beta_1\beta_2\,K^K,\qquad \eta=\int
dp_0\,\chi\,(p_0)\label{39}\end{equation} and the first-order energy shift is still given by (\ref{35}).
As for the propagator-approximated reduction, we can represent the various contributions by Feynman graphs. We can indeed get
$\,K^K\,$  by considering the transition operator
\begin{equation}T\,=\,(K^0+K^R\,)\,+\,(K^0+K^R\,)\,G^0\,(K^0+K^R\,)\,+\,\cdots\label{40}\end{equation}
and skipping some terms according to the two rules:\par
-- Each term must begin and end with $\,K^0$\par
-- There must be at least one $\,K^R$ between two consecutive $\,K^0.$
\par\noindent 
The 3D transition operator is now $\,-2i\pi\beta_1\beta_2T^K\,$ with 
$$T^K=K^K+K^KG^0K^K+\cdots=K^K(1-G^0K^K)^{-1}$$
$$=K^0(1-G^{KR}K^0)^{-1}(1-G^0K^0(1-G^{KR}K^0)^{-1})^{-1}$$
$$=K^0(1-G^{KR}K^0-G^0K^0)^{-1}=K^0(1-G^KK^0)^{-1}$$
$$=K^0(1-(1-G^0K^R\,)^{-1}G^0K^0\,)^{-1}$$
$$=K^0(1-G^0K^R-G^0K^0\,)^{-1}(1-G^0K^R\,)=K^0(1-G^0K\,)^{-1}(1-G^0K^R\,)$$
\begin{equation}K^0(1+G^0T\,)\,(1-G^0K^R\,)\label{41}\end{equation}
Replacing then $\,K^R\,$ by $\,K\!-\!K^0\,$ and $\,T(1\!-\!G^0K)\,$ by $\,K\,$ gives
\begin{equation}T^K\,=\,K^0+K^0G^0K^0+K^0G^0\,T\,G^0K^0.\label{42}\end{equation}
In contrast with what happens in the propagator-approximated reductions, the on mass shell restriction of this operator is
not the physical scattering amplitude of
field theory. In fact, the Bethe-Salpeter equation (\ref{1}), without inhomogeneous term, is a bound state equation, and for the
bound states we are only interested in the poles of (\ref{42}). These poles are the poles of $\,T,\,$ unless the integrations on
$\,p'_0,p_0\,$ make some residues vanish. Getting also the correct physical scattering amplitudes was a welcome bonus of the
propagator-approximated reductions.\par  If we want
$\,T\,$ in terms of
$\,T^K,\,$ we can write (\ref{41}) as
$$T^K(1-G^0K^R\,)^{-1}\,=\,K^0(1+G^0T\,)\,=\,(K-K^R\,)(1+G^0T\,)\, $$
\begin{equation}=\,T-K^R\,(1+G^0T\,)\,=\,(1-K^R\,G^0\,)T\,-K^R\label{42b}\end{equation} 
$$T\,=\,(1-K^R\,G^0\,)^{-1}\,T^K\,(1-G^0K^R\,)^{-1}\,+\,(1-K^R\,G^0\,)^{-1}\,K^R$$
\begin{equation}=\,(1+T^RG^0)\,T^K\,(1+G^0T^R)\,+\,T^R\label{43}\end{equation}
with
\begin{equation}T^R\,=\,K^R+K^RG^0K^R+\cdots.\label{44}\end{equation}
We have a large freedom in the choice of $\,K^0.\,$ We could choose
\begin{equation}K^0\,=\,\beta_1\beta_2\,\Lambda^{++}\,\beta_1\beta_2\,K(s,s)\,\Lambda^{++}.\label{45}\end{equation}
We could also keep only the ladder term of $\,K,\,$ or a part of it in (\ref{45}), like the Coulomb part of the one-photon exchange
contribution. We could use $\,K^T\,$ (in this case we expect that the remainder of $\,K^K\,$  does not finally contribute to the
total energy spectrum) or an approximation of it (one and two photon exchange terms, for example). Phillips and Wallace \cite{20}
suggested to choose $\,K^0\,$ in order to annihilate $\,g^{KR}\,$ exactly or up to a given order. If we choose to annihilate it at
first-order, the first-order energy shift (\ref{35}) vanishes and we get
\begin{equation}K^0={-1\over(2\pi)^2}\,\beta_1\beta_2\,(g^0)^{-1}\Lambda^{++}\int
dp'_0\,dp_0\,G^0(p'_0)K(p'_0,p_0)G^0(p_0)\,\beta_1\beta_2\,(g^0)^{-1}\Lambda^{++}.\label{46}\end{equation}
The skipping of terms in (\ref{40}) is then improved, as we must now have at last {\it two} $\,K^R$ between two consecutive
$\,K^0.\,$ If we choose to annihilate $\,g^{KR}\,$ exactly, we get, using (\ref{43}), (\ref{34}) and the relation
$\,G^{KR}\!=G^0T^RG^0:\,$
\begin{equation}{-1\over(2\pi)^2}\,\beta_1\beta_2\,(g^0)^{-1}\Lambda^{++}\int
dp'_0\,dp_0\,G^0(p'_0)T(p'_0,p_0)G^0(p_0)\,\beta_1\beta_2\,(g^0)^{-1}\Lambda^{++}=T^K\label{46c}\end{equation}
which is the $\,\Lambda^{++}\,$ projection of the expression given in \cite{20} (they use Salpeter's propagator and write thus
$\,\Lambda^{++}\!-\!\Lambda^{--}\,$ instead of $\,\Lambda^{++}\,$). In this case, $\,T^K\,$ gives the correct physical
scattering amplitudes when $\,P^0\!\to\! E_1\!+\!E_2,\,$ as the operators $\,(g^0)^{-1}\,$ kill the contributions of the
singularities other than the poles of $\,G^0.$         
 
\subsection{One-photon exchange.}   
Let us now compute (\ref{46}) with $\,K\,$ given by the one-photon exchange graph, in Feynman gauge:
\begin{equation}K\,=\,{2ie^2\over(2\pi)^3}\,\,\,{\gamma_1\!\cdot\gamma_2\over(p'\!-\!p)^2+i\epsilon}\label{47}\end{equation}
Writing $\,p'\!-\!p\!=\!k\,$ and leaving out trivial factors, we have to compute the integral
$$I=\int dp_{10}\,dk_0\quad{1\over p_{10}+k_0-E'_1+i\epsilon}\quad{1\over P_0-p_{10}-k_0-E'_2+i\epsilon}\,\,$$
\begin{equation}{1\over
k_0^2-\vec k^2+i\epsilon}\quad{1\over p_{10}-E_1+i\epsilon}\quad{1\over P_0-p_{10}-E_2+i\epsilon}\label{48}\end{equation}
Performing the integration in $\,p_{10}\,$ by closing the integration path clockwise, we get the residues of the poles at 
$\,p_{10}\!=\!E'_1\!-\!k_0,\,E_1:$
$$I=-2i\pi\int_{(\eta)} \!dk_0\quad{1\over
k_0^2-\vec k^2+i\epsilon}\quad\left[\quad{1\over P_0-E'_1-E'_2+i\epsilon}\quad{1\over E'_1-E_1-k_0+i\epsilon}\right.$$
$${1\over P_0-E'_1-E_2+k_0+i\epsilon}\quad +\quad
{1\over E_1-E'_1+k_0+i\epsilon}\quad{1\over P_0-E_1-E'_2-k_0+i\epsilon}$$
\begin{equation}\left.{1\over P_0-E_1-E_2+i\epsilon}\quad\right]\label{49}\end{equation}
The possibility of $\,k_0\,$ being equal to $\,E'_1\!-\!E_1,\,$ leading to a double pole, has been taken into account by leaving an
interval of length $\,\eta\,$ around this value in order to take the $\,\eta\!\to\!0\,$ limit afterwards. This can also be done by
replacing the poles $\,\pm(E'_1\!-\!E_1-k_0\pm i\epsilon)^{-1}\,$ by their principal parts, or equivalently by adding a
$\,i\pi\delta(k_0-E'_1+E_1)\,$ to each one.  For the integral in
$\,k_0\,$ we shall close the integration path counterclockwise in the first term and clockwise in the second one. In each case we
shall have a pole of the propagator at
$\,k_0\!=\!E'_1\!-\!E_1\,$ (the $\,\delta\,$ reduces the contribution of this pole by a factor 2)  and a pole of the kernel at
$\,k_0\!=\!-\vert\vec k\vert\,$ and $\,k_0\!=\!\vert\vec k\vert\,$ respectively. The final result is:
$$I=-4\pi^2\,\, {1\over P_0-E'_1-E'_2}\quad{1\over (E'_1-E_1)^2-\vec k^2}\quad{1\over P_0-E_1-E_2}\,\,\bigg[\,1\,+$$    
\begin{equation}\left. {1\over2\vert\vec k\vert}\left\{{(E'_1-E_1-\vert\vec k\vert\,)(P_0-E_1-E_2)\over
P_0-E'_1-E_2-\vert\vec k\vert\,}\,-\,{(E'_1-E_1+\vert\vec k\vert\,)(P_0-E'_1-E'_2)\over
P_0-E_1-E'_2-\vert\vec k\vert\,}\right\}\right]\label{50}\end{equation}Here again, we must take the principal parts of the poles in
$\,\vert\vec k\vert\!=\!\vert E'_1\!-\!E_1\vert\,$ in the future integrals in $\,\vec k.\,$ The first term comes from the residues of
the poles of the propagators only, while the two other terms combine the residues of one pole of the propagators and one pole of the
kernel. Closing the integration paths in the opposite directions would interchange the two fermions in the first term and modify the
two other terms accordingly. At lowest order, the expression between brackets becomes
\begin{equation}[\cdots]\,\approx\,1\,+\,{1\over2\vert\vec k\vert}\,(2P_0-E'_1-E'_2-E_1-E_2).\label{51}\end{equation}
We could define $\,K^0\,$ as the kernel which leads to (\ref{50}), to the first term of (\ref{50}), or to the first term of
(\ref{50}) with $\,(E'_1\!-\!E_1)^2\!-\!\vec k^2\,$ replaced by simply $\,-\!\vec k^2\,$ (Coulomb potential). We see anyway
that the relative time dependence of the kernel contibutes more to (\ref{50}) via the residues of its poles than directly. We
should however consider also the contribution of the lowest-order crossed graph, which could partially cancel (\ref{50}). We
know indeed that, in well chosen propagator-approximated reductions, all higher-order contributions cancel mutually at the
one-body limit \cite{15,19}.
      
\section{The three-fermion problem.}

\subsection{Three-fermion Bethe-Salpeter equation.}
 The Bethe-Salpeter equation is now:
\begin{equation}\Phi=\left[G^0_1G^0_2K_{12}+G^0_2G^0_3K_{23}+G^0_3G^0_1K_{31}+G^0_1G^0_2G^0_3K_{123}\right]
\Phi\label{52}\end{equation} 
where $K_{123}$ is the sum of the purely three-body irreducible contributions. Let us first consider the case of
three two-body positive-energy, instantaneous and cluster-energy independent interactions $\,K^0_{ij}\,$ (i.e. containing
$\,\Lambda^{++}_{ij}\,$ projectors  and independent of
$\,p'_{ij0},p_{ij0},P_{ij0}\,$).  The Bethe-Salpeter equation becomes in this case
\begin{equation}\Phi=\,{-1\over2i\pi}\,G^0_1G^0_2G^0_3\,\beta_1\beta_2\beta_3\,\left[\,V_{12}\,\psi_{12}\,+\,V_{23}
\,\psi_{23}\,+\,V_{31}\,\psi_{31}\,\right]\label{53}\end{equation}
\begin{equation}V_{ij}\,=\,-2i\pi\,\beta_i\beta_jK^0_{ij},\qquad \psi_{ij}(p_{k0})=\beta_k\,G_{0k}^{-1}\int
dp_{ij0}\,\Phi.\label{54}\end{equation} This leads to a set of three coupled integral equations in the $\,\psi_{ij}.\,$  
We shall search for solutions analytical in the Im($p_{k0})\!<\!0\,$ half  planes and close the integration paths
clockwise in these planes. The only singularities will then  be the poles of the free propagators. Performing the
integrations with respect to the $\,p_{ij0}\,$
 gives then
\begin{equation}\psi_{12}(p_{30})=\,{\Lambda^{++}_{12}\over
(P_0-S)-(p_{30}-h_3)+i\epsilon}\,\left[\,V_{12}\psi_{12}(p_{30})
+V_{23}\psi_{23}(h_1)+V_{31}\psi_{31}(h_2)\,\right]\label{55}\end{equation} and similarly for $\,\psi_{23}\,$ and
$\,\psi_{31}\,$ (from now on, we shall often write the equations for the (12) pair only, understanding the two other equations
obtained by circular permutations on the $\,ijk\,$ indexes). Solving (\ref{55}) with respect to
$\,\psi_{12}(p_{30})\,$ gives a wave function which is analytical in the Im($p_{30})\!<\!0\,$ half plane, and confirms thus the
existence of such solutions. Furthermore, equation (\ref{55}) shows that the three projections
$\,\Lambda^+_k\psi_{ij}(h_k)\,$ are equal  (let us call them $\,\psi\,$) and satisfy the 3D equation
\begin{equation}\psi={\Lambda^{+++}\over
P_0-E_1-E_2-E_3}\,\left[\,V_{12}+V_{23}+V_{31}\,\right]\,\psi.\label{56}\end{equation}
Moreover,  $\,\psi\,$ is the integral of the Bethe-Salpeter amplitude with respect to the relative
times:         
\begin{equation}\psi={-1\over2i\pi}\,\int dp_0\,\Phi\equiv{-1\over2i\pi}\,\int
dp_{10}dp_{20}dp_{30}\,\delta(p_{10}+p_{20}+p_{30}-P_0)\,\Phi.\label{57}\end{equation}
In \cite{21}, we proposed the choice
\begin{equation}K^0_{ij}(\,p'_{ij0},\,p_{ij0},\,P_{ij0})\,=\beta_i\beta_j\,\Lambda^{++}_{ij}\,\beta_i\beta_j
\,K^T_{ij}(\,s_{ij},\,s_{ij},\,P_0-h_k\,)\,\Lambda^{++}_{ij}\label{58}\end{equation}
in order to recover, at the three 2+1 separated clusters limits, the propagator-approximated 3D reductions (subsection 2.2
above).\par For the general three-fermion Bethe-Salpeter equation (\ref{52}) it seemed to be a good idea to try an expansion around
a set of positive-energy instantaneous approximations of the two-body kernels. We first tried to start with approximations giving
the 3D equation (\ref{56}) in a more direct way than above, like
\begin{equation}G^0_1G^0_2\,\approx\,G^{\delta}_{12}\label{59}\end{equation}
\begin{equation}G^0_1G^0_2\,\delta(p'_{30}-p_{30})\,\approx\,G^{\delta}_{12}\,\delta(p'_{30}-h_3)\label{60}\end{equation}
\begin{equation}K_{12}\,\delta(p'_{30}-p_{30})\,\approx\,\Lambda^+_3\,
K^0_{12}\,\delta(p'_{30}-h_3)\label{61}\end{equation}
with $\,K^0_{12}\,$ given by (\ref{58}). The zero-order 3D equation is in each case (\ref{56}), but the correction terms
spoil the result: with (\ref{59}) or (\ref{60}) they contain supplementary zero-order contributions. With (\ref{61}) they
contain no supplementary zero-order contributions, but the correction terms can not easily be sorted by increasing
order. We must thus proceed more carefully. We finally decided to start with a positive-energy, instantaneous and cluster-energy
independent approximations of the two-body kernels, like (\ref{58}) or its Born approximation, adding the third $\,\Lambda^+_k\,$
projector in order to avoid unnecessary complications, and to proceed in three steps, inspired by the demonstration
(\ref{52})-(\ref{56}):\par
i) Transform the Bethe-Salpeter equation (\ref{52}) into a set of three coupled equations for the $\,\psi_{ij}(p_{k0}).$\par
ii) Transform this set into a set of three coupled 3D equations for the $\,\psi_{ij}(h_k)\,$ (these are {\it not} the Faddeev
equations).\par iii) Transform this last set into a single 3D equation for $\,\psi,\,$ defined as the mean of the
$\,\psi_{ij}(h_k).$\par         

\subsection{Three coupled equations for three cluster-energy depending amplitudes.}
Let us first write the Bethe-Salpeter equation (\ref{52}) in a more compact form:
\begin{equation}\Phi\,=\,G^0K\,\Phi\label{62}\end{equation}
with
\begin{equation}G^0\,=\,G^0_1\,G^0_2\,G^0_3\label{63}\end{equation}
\begin{equation}K\,=\,(G^0_3)^{-1}K_{12}\,+\,(G^0_1)^{-1}K_{23}\,+\,(G^0_2)^{-1}K_{31}\,+\,K_{123}.\label{64}\end{equation}   
With these notations, we can formally perform the same manipulations as in the two-fermion problem (subsection 2.3): 
\begin{equation}K\,=\,K^0\,+\,K^R\label{65}\end{equation}
\begin{equation}\Phi=G^0K^0\Phi+G^0K^R\Phi\label{66}\end{equation}  
\begin{equation}\Phi=(1-G^0K^R)^{-1}G^0K^0\Phi=G^K\,K^0\Phi\label{67}\end{equation}
with
\begin{equation}G^K\,=\,G^0+G^0K^RG^0+G^0K^RG^0K^RG^0+\cdots\equiv\,G^0+G^{KR}.\label{68}\end{equation}
Since eq. (\ref{67}) ends with $\,K^0\Phi,\,$ we can transform it into a set of three coupled equations for the
$\,\psi_{ij}(p_{k0}),\,$ using the definition (\ref{54}):
\begin{equation}\psi_{12}(p'_{30})\,=\,\sum_k\int
dp_{k0}\,G^K_{12,ij}(p'_{30},p_{k0})\,K^0_{ij}\,\psi_{ij}(p_{k0})\label{69}\end{equation} 
with
\begin{equation}G^K_{12,ij}(p'_{30},p_{k0})\,=\,\beta_3\,\big[G^0_3(p'_{30})\big]^{-1}\int
dp'_{120}\,dp_{ij0}\,G^K(p'_0,p_0)\,\,\beta_k\label{70}\end{equation}

\subsection{Three coupled 3D equations.}
Let us now compute the  contribution of $\,G^0\,$ to (\ref{70}). We get
\begin{equation}G^0_{12,12}(p'_{30},p_{30})\,=\,{-2i\pi\over (P_0-S)-(p'_{30}-h_3)+i\epsilon}\,
\beta_1\beta_2\,\delta(p'_{30}-p_{30})\label{72}\end{equation} 
\begin{equation}G^0_{12,23}(p'_{30},p_{10})\,=\,\beta_3\,G^0_1(p_{10})\,G^0_2(P_0-p_{10}-p'_{30})\,\beta_1\label{73}\end{equation}   
\begin{equation}G^0_{12,31}(p'_{30},p_{20})\,=\,\beta_3\,G^0_2(p_{20})\,G^0_1(P_0-p_{20}-p'_{30})\,\beta_2\label{74}\end{equation}
The product of propagators (\ref{73})  will be included in a clockwise path integral in
$\,p_{10},\,$ to which only the pole of $\,G^0_1(p_{10})\,$ at $\,p_{10}\!=\!h_1\,$ contributes. This is no more true in the general
case, but we can isolate a part of $\,G^0_1(p_{10})\,$ which leads to the same result:
\begin{equation}G^0_1(p_{10})\,=\,G^{\delta}_1(p_{10})\,+\,G^R_1(p_{10})\label{75}\end{equation}
with
\begin{equation}G^{\delta}_1(p_{10})\,=\,-2i\pi\,\delta(p_{10}-h_1)\,\beta_1\label{76}\end{equation}
\begin{equation}G^R_1(p_{10})\,=\,{1\over p_{10}-h_1-i\epsilon}\,\,\beta_1\label{77}\end{equation}
so that
\begin{equation}G^{\delta}_{12,23}(p'_{30},p_{10})\,=\,{-2i\pi\over (P_0-S)-(p'_{30}-h_3)+i\epsilon}\,
\beta_2\beta_3\,\delta(p_{10}-h_1)\label{78}\end{equation}
and similarly 
\begin{equation}G^{\delta}_{12,31}(p'_{30},p_{20})\,=\,{-2i\pi\over (P_0-S)-(p'_{30}-h_3)+i\epsilon}\,
\beta_3\beta_1\,\delta(p_{20}-h_2).\label{79}\end{equation}
Putting all together, we have finally
$$\psi_{12}(p'_{30})\,=\,{-2i\pi\over (P_0-S)-(p'_{30}-h_3)+i\epsilon}\,
\beta_1\beta_2\,K^0_{12}\,\psi_{12}(p'_{30})$$ 
$$+\,{-2i\pi\over (P_0-S)-(p'_{30}-h_3)+i\epsilon}\,
\big[\,\beta_2\beta_3\,K^0_{23}\,\psi_{23}(h_1)\,+\,\beta_3\beta_1\,K^0_{31}\,\psi_{31}(h_2)\,\big]$$          
$$+\,\int dp_{10}\,\beta_3\,G^R_1(p_{10})\,G^0_2(P_0-p_{10}-p'_{30})\,\beta_1\,K^0_{23}\,\psi_{23}(p_{10})$$
$$+\,\int dp_{20}\,\beta_3\,G^R_2(p_{20})\,G^0_1(P_0-p_{20}-p'_{30})\,\beta_2\,K^0_{31}\,\psi_{31}(p_{20})$$
\begin{equation}+\,\sum_k\int
dp_{k0}\,G^{KR}_{12,ij}(p'_{30},p_{k0})\,K^0_{ij}\,\psi_{ij}(p_{k0})\label{80}\end{equation}
If we neglect the last term of (\ref{80}) the two preceding terms vanish also by integration, and we recover the result of the
positive-energy instantaneous approximation. We shall thus move the first term of (\ref{80}) to the left-hand side, consider
the two next terms (second line) as the principal contributions, and the three last terms as perturbations. The difference
between this approach and that based on (\ref{61}) lies in the fact that we do not replace immediately $\,p'_{30}\,$ by
$\,h_3\,$ in the first term. We get
$$\psi_{12}(p'_{30})\,=\,g_{12}(p'_{30})\,\big[\,V_{23}\,\psi_{23}(h_1)\,+\,V_{31}\,\psi_{31}(h_2)\,\big]$$
\begin{equation}+\,\,g_{12}(p'_{30})\,[\,g^0_{12}(p'_{30})\,]^{-1}\,\sum_k\int
dp_{k0}\,G^{KT}_{12,ij}(p'_{30},p_{k0})\,K^0_{ij}\,\psi_{ij}(p_{k0})\label{81}\end{equation}
where $\,G^{KT}_{12,ij}\,$ is defined from $\,G^{KR}_{12,ij}\,$ by including the 3th and the 4th lines of (\ref{80}) in
the last one, and 
\begin{equation}g_{12}(p'_{30})\,=\,{1\over (P_0-S-V_{12})-(p'_{30}-h_3)+i\epsilon}\label{82}\end{equation}
\begin{equation}g^0_{12}(p'_{30})\,=\,{1\over (P_0-S)-(p'_{30}-h_3)+i\epsilon}\label{83}\end{equation}  
 The iterations of (\ref{81}) lead to
$$\psi_{12}(p'_{30})\,=\,g_{12}(p'_{30})\,\big[\,V_{23}\,\psi_{23}(h_1)\,+\,V_{31}\,\psi_{31}(h_2)\,\big]$$
$$+\,\,g_{12}(p'_{30})\,[\,g^0_{12}(p'_{30})\,]^{-1}\,\sum_k\int
dp_{k0}\,G^{\,TT}_{12,ij}(p'_{30},p_{k0})\,K^0_{ij}$$
\begin{equation}g_{ij}(p_{k0})\,\big[\,V_{jk}\,\psi_{jk}(h_i)\,+\,V_{ki}\,\psi_{ki}(h_j)\,\big].\label{84}\end{equation}
defining $\,G^{TT}\,$ as the sum of the iterations of $\,G^{KT}.$\par 
We can now take (\ref{84}) at $\,p'_{30}\!=\!h_3,\,$ multiply at left by $\,[\,g_{12}(h_3)]^{-1},\,$  and move the term in
$\,V_{12}\psi_{12}(h_3)\,$ back to the right-hand side:   
$$\psi_{12}(h_3)\,=\,{1\over P_0-S}\,\big[\,V_{12}\,\psi_{12}(h_3)\,+\,V_{23}\,\psi_{23}(h_1)\,+\,V_{31}\,\psi_{31}(h_2)\,\big]$$
\begin{equation}+\,\sum_k\,M_{12,ij}\,\big[\,V_{jk}\,\psi_{jk}(h_i)\,+\,V_{ki}\,\psi_{ki}(h_j)\,\big]\label{85}\end{equation}
with
\begin{equation}M_{12,ij}\,=\,\int
dp_{k0}\,G^{\,TT}_{12,ij}(h_3,p_{k0})\,K^0_{ij}\,g_{ij}(p_{k0}).\label{86}\end{equation} 

\subsection{One 3D equation.}
Let us write (\ref{85}) and the two other equations obtained by circular permutations in matricial form. We shall define
\begin{equation}\Psi^k\,=\,\psi_{ij}(h_k),\qquad \Pi^{ij}\,=\,{1\over3},\quad \overline\Pi \,=\,1-\Pi.\label{87}\end{equation}
The matrix $\,\Pi\,$ is the projector ($\Pi^2\!=\!\Pi\,$) on the $\,u^k\!=\!1\,$ vector (of norm 3). We have
\begin{equation}\Pi\,\Psi\,=\,\psi\,u\,,\qquad \psi\,=\,{1\over3}\,(\,\Psi^1\,+\,\Psi^2\,+\,\Psi^3\,).\label{88}\end{equation}
We can write eq. (\ref{85}) in the form
\begin{equation}
\Psi\,=\,M\,\Psi,\quad M=M^0+M^R,\quad M^{0i1}\,=\,{1\over P_0-S}\, V_{23},\,\, etc...\label{89}\end{equation}
The action of the projector $\,\Pi\,$ on $\,M^0\,$ is
\begin{equation}\Pi\,M^0\,=\,M^0,\quad M^0\,\Pi\,=\,{\Pi\over P_0-S}\,V,\quad V=V_{12}+V_{23}+V_{31}.\label{90}\end{equation}
We shall now transform the matricial equation (\ref{89}) into a scalar equation for $\,\psi.\,$ The expansion of (\ref{89})
with respect to $\,\Pi\Psi\,$ gives
\begin{equation}\Pi\,\Psi\,=\,\Pi\,M\,(\,1+\overline\Pi M+\overline\Pi M\,\overline\Pi
M+\cdots)\,\Pi\,\Psi\label{91}\end{equation}
in which $\,\overline\Pi M\!=\!\overline\Pi M^R.\,$ Taking the scalar product with $\,u\,$ leads finally to
\begin{equation}\psi\,=\,\left[\,{1\over P_0-S}\,V\,+\,{1\over3}\,u^{\tau}\,(\,M^R+M\,[\,\overline\Pi M^R+\overline\Pi
M^R\,\overline\Pi M^R+\cdots]\,)\,u\,\right]\,\psi.\label{92}\end{equation}
In this equation, $\,V\,$ contains the projector $\,\Lambda^{+++}\,$ at left and at right, while $\,M^R\,$ contains it at
right. We can thus trivially transform (\ref{92}) into an equation for $\,\Lambda^{+++}\psi.\,$            
             
\subsection{First-order energy shift.}
The 3D reduction above is rather complicated, since it implies three levels of series expansions. The computation of the first-order
energy shift remains however tractable. The energy shift is, keeping only the terms which contribute at first-order:
\begin{equation}P_0-P_0^{(0)}\,=\,<(P_0-S)\,{1\over3}\,u^{\tau}\,(\,1+M^0\overline\Pi\,)\,M^R\,u>.\label{93}\end{equation}
We can make the replacements
\begin{equation}1+M^0\overline\Pi\,=\,1+M^0-{\Pi\over P_0-S}\,V\,\to\,M^0\label{94}\end{equation}
as the contributions of the first and third terms cancel mutually. We remain with
$$P_0-P_0^{(0)}\,=\,<(P_0-S)\,{1\over3}\,u^{\tau}\,M^0\,M^R\,u>$$
\begin{equation}=\,\sum_{k',k}<V_{i'j'}\,M_{i'j',ij}\,(\,V-V_{ij}\,)>\label{95}\end{equation} 
The contribution of the terms with $\,i'j'\!=\!12\,$ is then
\begin{equation}\Delta_{12}\,=\,<V_{12}\,\sum_k\int
dp_{k0}\,G^{\,TT}_{12,ij}(h_3,p_{k0})\,K^0_{ij}\,g_{ij}(p_{k0})\,(\,V-V_{ij}\,)>\label{96}\end{equation}
At first order, we can replace $\,G^{TT}_{12,ij}\,$ by the non-iterated $\,G^{KT}_{12,ij}\,$ and further by $\,G^{KR}_{12,ij},\,$ as
the 3th and 4th lines of (\ref{80}) will lead to null integrals. Using the definitions (\ref{68}), (\ref{70}) we
get
$$\Delta_{12}\,=\,<V_{12}\,\sum_k\int\delta(p'_{30}-h_3)
dp'_0\,dp_0\,\beta_3\,\big[G^0_3(p'_{30})\big]^{-1}$$
\begin{equation}G^{KR}(p'_0,p_0)\,\,\beta_k\,K^0_{ij}\,g_{ij}(p_{k0})\,(\,V-V_{ij}\,)>\label{97}\end{equation}\vskip5mm\noindent
Replacing $\,V\,$ by $\,P_0\!-\!S\,$ and $\,G^{KR}\,$ by $\,G^0K^RG^0\,$ leads to
\begin{equation}P_0-P_0^{(0)}\,=\,{-1\over4\pi^2}<\int
dp'_0\,dp_0\,V_F(p'_0)\,G^0(p'_0)K^R(p'_0,p_0)G^0(p_0)\,\beta_1\beta_2\beta_3\,V_I(p_0)>\label{98}\end{equation}         
with
\begin{equation}V_I(p_0)\,=\,\sum_k\,V_{ij}
\,g_{ij}(p_{k0})\,g_{ij}^{-1}(h_k)\label{99}\end{equation}
\begin{equation}V_F(p'_0)\,=-2i\pi\,\sum_{k'}\,V_{i'j'}\,\delta(p'_{k'0}-h_{k'})
\,\beta_{k'}\,\big[G^0_{k'}(p'_{k'0})\big]^{-1}.\label{100}\end{equation}
In writing (\ref{100}), we mean that $\,\big[G^0_{k'}(p'_{k'0})\big]^{-1}$ cancels the  $\,G^0_{k'}(p'_{k'0})\,$  contained in
$\,G^0(p'_0)\,$ before the replacement of $\,p'_{k'0}\,$ by $\,h_{k'}.\,$     The energy shift (\ref{98}) contains 36 terms: there
are 4 terms in
$\,K^R\,$ (if we do not forget the irreducible three-body term) times 3 terms in $\,V_I\,$ and 3 terms in $\,V_F\,$.  Our reason for
introducing these
$\,V_I\,$ and $\,V_F\,$ is the fact that they both could be replaced by $\,V\,$  if $\,K^R\,$ were instantaneous.\par
The singularities to consider when performing the integrations are the poles of the propagators, the singularities of
$\,K^R\,$ and the poles of $\,V_I.\,$ It is always possible to choose the
integration paths, in each term, in order to avoid contributions from the poles of $\,V_I.\,$  The contributions of the poles of the
propagators are then easy to compute, as  $\,g_{ij}(p_{k0})\,$ in (\ref{99}) is to be taken at 
$\,p_{k0}\!=\!h_k.\,$ We get 
\begin{equation}R_{12}\,=\,<\sum_{k'k}\,V_{i'j'}\,{1\over P_0-S}\,V^{Rba}_{12}\,{1\over
P_0-S}\,V_{ij}>\label{101}\end{equation}
plus the similar $\,R_{23}\,$ and $\,R_{31}\,$ and a contribution $\,R_{123}\,$ of the irreducible three-body kernel.  By
$\,V^{Rba}_{12}\,$ we denote
$\,-2i\pi\beta_1\beta_2K^R_{12}\,$ with the initial fermion a, the final fermion b  and the spectator fermion 3 on their
positive-energy mass shells. We have to consider 9 terms in 4 different groups:\hfill\break\par
$\,k'=k=3\,$\qquad\qquad\qquad  $\,ba=11,\,12,\,21\,$ or $\,22$\par
$\,k'=1\,\hbox{or}\,2,\,k=3\,$\qquad\qquad  $\,ba=k'1\,$ or $\,k'2$\par
$\,k'=3,\,k=1\,\hbox{or}\,2\,$\qquad \qquad $\,ba=1\,k\,$ or $\,2\,k$\par
$\,k'=1\,\hbox{or}\,2\,,\,k=1\,\hbox{or}\,2\,$\qquad\qquad  $\,ba=k'k\,$\par\hfill\break
We still have some freedom in the choice of $\,ba\,$ which we could use to get a real energy shift, for example by choosing
$\,a\!=\!b\,$ in the three first groups. The contributions of an instantaneous part $\,K^{RI}_{12}\,$ of $\,K^{R}_{12}\,$
would sum into  a $\,<V^{RI}_{12}>\,$ but we could also include this $\,K^{RI}_{12}\,$ directly in $\,K^0_{12}.\,$ The corrections
corresponding to the differences between the various $\,V^{Rba}_{12}\,$ are a step towards Gross' spectator model
\cite{22,23}.\par The principal task remains however the calculation of the contributions of the singularities of the
$\,K^R_{ij},\,$ for the integration paths defined by our choices of $\,ab\,$ above. As we saw in section 2.4, these contributions
could be as important as these coming directly from the relative time dependence of these kernels (they could however be partially
cancelled by the contributions of two-body non-ladder terms and three-body terms \cite{15,19,22} ). A general expression would be
obtained by removing the delta part of the propagators containing the poles considered in (\ref{101}), using (\ref{70}).  To go
further, however, we should specialize to specific three-fermion problems. It must be noted that the ratio
$\,g_{ij}(p_{k0})\,g_{ij}^{-1}(h_k)\,$ of (\ref{99}) is no more 1 in the terms containing $\,G^R_k(p_{k0}).\,$ If the singularities
of
$\,K^R\,$ in the complex $\,p_{k0}\,$ plane lie far from the positive-energy mass shell, this ratio can be approximated by 
$\,-(p_{k0}\!-\!h_k)^{-1}\,g_{ij}^{-1}(h_k)\,$ and takes a relatively low value, suggesting us to neglect the contribution of the
corresponding $\,G^R_k(p_{k0})\,$ to (\ref{98}). We could then make the replacement
\begin{equation}V_I(p_0)\to\,=-2i\pi\,\sum_{k\neq u}\,V_{ij}\,\beta_k\,\big[G^0_k(p_{k0})\big]^{-1}\delta(p_{k0}-h_k)\,
+\,\sum_{k= u}\,V_{ij}\label{104}\end{equation}
in (\ref{98}). We isolated the contributions of the unconnected terms $\,k\!=\!u\,$ (like $\,k\!=\!3\,$ combined with
$\,k'\!=\!3\,$ and $\,K^R_{12}\,$). The resulting energy shift is now manifestly real.         

\subsection{Cluster separability.}
Physically, if we "switch off" the (23), (31) and (123) interactions, we are left with a pair of interacting
fermions (12) and a free fermion (3). We would like to read this on our final 3D equation. however:\par
-- The cluster-separability condition may be satisfied by an exact 3D reduction, but not by its truncated
forms.\par
-- Even with an exact 3D reduction, the cluster separability property is a welcome bonus, not a requirement: the
Bethe-Salpeter equation (\ref{52}), from which we started, is indeed correct for three-fermion bound states
only.\par
In our calculations above, the cluster-separability property is in general spoiled by our iteration of
(\ref{81}), which starts with a term which vanishes at the (12) cluster-separated limit. The (12)-limit 
equation is then built with the approached $\,-2i\pi\beta_1\beta_2 K^0_{12}\,$ only. The cluster separability
can however be restored by performing a supplementary iteration, like our replacement of $\,V\,$ by
$\,P_0\!-\!S\,$ in (\ref{95}), leading the the cluster-separable correction (\ref{101}). Another
possibility consists in our previous choice (\ref{58}) of $\,K^0_{ij}.\,$ With this choice, the
cluster-separated limits are exact. Moreover, the corresponding two-body scattering amplitudes are also
correct. Our choice (\ref{58}) of $\,K^0_{ij}\,$ is thus a priori the most interesting. A simpler
choice (as its Born approximation) would however not be uncorrect in the calculation of the
higher-order corrections to the three-fermion bound state spectrum.                              

\section{Conclusions}
The kernel-approximated reduction is an interesting alternative to the more usual propagator-approximated reduction, although the
resulting 3D equation is valid only for bound states. In both cases, the 3D potentials can be obtained with simple modifications of
the Feynman rules which give the transition operator in field theory. Further investigations on the kernel-approximated
reductions, with comparison with the other method (one-body limit, introduction of an external potential, gauge invariance, numerical
examples, etc...) would be useful.\par Our 3D reduction of the three-fermion Bethe-Salpeter equation provides in principle a way of
progressively improving a manageable approximated 3D equation. We use it to give a manageable expression of the first-order energy
shift. The contributions of the free propagator poles to the energy shift are easily calculated in terms of the Bethe-Salpeter
kernels with two of the fermions on their mass shell, in the spirit of Gross' spectator model. The contributions of the
singularities of the kernels themselves should be computed in more specific frameworks ($\,He^3\,$ atom, triton, three-quark
systems...) and could be important, as suggested by our computation of the first-order energy shift in the framework of the
one-photon exchange model for the two-fermion problem.


\begin{thebibliography}{99}
\bibitem{1} E.E. Salpeter and H.A. Bethe: { Phys. Rev.} {\bf 84}  1232 (1951) 
\bibitem{2} M. Gell-Mann and F. Low: { Phys. Rev.} {\bf 84}  350 (1951)
\bibitem{3} E.E. Salpeter: { Phys. Rev.} {\bf 87}  328 (1952)
\bibitem{4} A.A. Logunov and A.N. Tavkhelitze: { Nuovo Cimento} {\bf 29}  380 (1963)
\bibitem{5} R. Blankenbecler and R. Sugar: {Phys. Rev.} {\bf 142}  1051 (1966)
\bibitem{6} I.T. Todorov: { Phys. Rev.} {\bf D3}  2351 (1971)
\bibitem{7} C. Fronsdal and R.W. Huff: { Phys. Rev.} {\bf D3}  933 (1971)
\bibitem{8} A. Klein and T.S.H. Lee: { Phys. Rev.} {\bf D10}  4308 (1974)
\bibitem{9} G.P. Lepage: { Phys. Rev.} {\bf A16}  863 (1977);
 G.P. Lepage: Ph D dissertation, SLAC Report n 212 (1978); 
 W.E. Caswell and G.P. Lepage: { Phys. Rev.} {\bf A18}  810 (1978)
\bibitem{10} J.L. Gorelick and H. Grotch: { Journ. Phys. G} {\bf 3}  751 (1977)
\bibitem{11} G.T. Bodwin, D.R. Yennie and M.A. Gregorio: {Rev. Mod. Phys.} {\bf 57}  723 (1985)
\newline 
 J.R. Sapirstein and D.R. Yennie, in {\it Quantum Electrodynamics} edited by T. Kinoshita
(World Scientific, Singapore 1990) p.560. 
\bibitem{12} V.B. Mandelzweig and S.J. Wallace: { Phys. Lett.} {\bf B197} 469 (1987)\newline
   S.J. Wallace and V.B. Mandelzweig: { Nucl. Phys.} {\bf A503}  673 (1989)
\bibitem{13} E.D. Cooper and B.K. Jennings: { Nucl. Phys.} {\bf A483}  601 (1988)
\bibitem{14} J.H. Connell: { Phys. Rev.} {\bf D43}  1393 (1991)
\bibitem{15} F. Gross: { Phys. Rev.} {\bf 186}  1448 (1969); { Phys. Rev.} {\bf C26}  2203 (1982)
\bibitem{16}	H. Sazdjian: {  J. Math. Phys.} {\bf 28}   2618 (1987); {\bf 29}   1620 (1988)
\bibitem{17} J. Bijtebier and J. Broekaert: {   Nuovo Cimento} {\bf A105}  351, 625 (1992)
\bibitem{19} J. Bijtebier and J. Broekaert:  { Journal of Physics~G} {\bf 22}  559, 1727 (1996)
\bibitem{18} J. Bijtebier and J. Broekaert: { Nucl. Phys.} {\bf A612} 279 (1997)
\bibitem{20} D.R. Phillips and S.J. Wallace: Phys. Rev. {\bf C54} 507 (1996); Few-Body Systems {\bf 24} 175 (1998)
\bibitem{21} J. Bijtebier: A three-fermion Salpeter equation. Preprint VUB/TENA/98-02 (hep-th/9809131) (1998). Submitted to Few-Body
Systems
\bibitem{22} F. Gross: { Phys. Rev.} {\bf C26}  2226 (1982)
\bibitem{23} A. Stadler, F. Gross, M. Frank: { Phys. Rev.} {\bf C56}  2396 (1997)









\end{thebibliography}
\end{document}